\documentclass{article}

\usepackage[numbers]{natbib}
\usepackage[preprint]{neurips_2025}
\usepackage{hyperref}
\usepackage{tcolorbox}
\usepackage{booktabs}
\usepackage{graphicx}
\usepackage{colortbl}
\usepackage{fontawesome5}

\definecolor{secAcaColor}{HTML}{E8F1FA}
\definecolor{secIndColor}{HTML}{FBEEDC}
\definecolor{secConColor}{HTML}{E9F3E4}
\newcommand{\sectorAca}{\cellcolor{secAcaColor}\faGraduationCap~Academia}
\newcommand{\sectorInd}{\cellcolor{secIndColor}\faBuilding~Industry}
\newcommand{\sectorCon}{\cellcolor{secConColor}\faUserTie~Consultancy}
\newcommand{\rowh}{\rule{0pt}{2.6ex}}

\newtcolorbox{quotebox}[1][]{
  colback=gray!5,
  colframe=gray!60,
  fonttitle=\bfseries\small,
  title={#1},
  boxrule=0.5pt,
  arc=2pt,
  left=6pt,
  right=6pt,
  top=4pt,
  bottom=4pt,
  fontupper=\small\itshape
}

\newif\ifcomments
\commentstrue

\ifcomments
  \newcommand{\niklas}[1]{\textcolor{blue!70!black}{\textbf{[Niklas:} #1\textbf{]}}}
\else
  \newcommand{\niklas}[1]{}
\fi

\begin{document}

%% ---------------------------------------------------------------------
%% Title and Authors
%% ---------------------------------------------------------------------

\title{The Future of Visualization Dashboards in the Age of Generative AI}

\author{%
Vaishali Dhanoa~\thanks{Correspondence: \texttt{dhanoa@cs.au.dk}.}\textsuperscript{1,2}
\And Duosi Dai \textsuperscript{1}
\And Gabriela Molina Le\'on\textsuperscript{1}
\And Eduard Gr\"{o}ller\textsuperscript{2}{3}
\And Niklas Elmqvist~\textsuperscript{1}
\\[4pt]
\textsuperscript{1}\,Aarhus University \quad
\textsuperscript{2}\,TU Wien \quad
\textsuperscript{3}\,VRVIS GmbH 
}

\maketitle

%% ---------------------------------------------------------------------
%% Abstract
%% ---------------------------------------------------------------------
\begin{abstract}
  Generative AI promises easier dashboard creation, raising questions about the future of dashboards and the people who create and use them. We interviewed 16 experts based in 14 countries about their practices and expectations. Almost all expected dashboards to persist for recurring questions, monitoring, and reporting. They anticipated adaptive views and combinations of language, graphical controls, and gestures, while emphasizing interaction as part of human exploration and understanding. Participants expected authors' responsibilities to shift toward specifying requirements, curating generated work, and evaluating outputs, with design knowledge and communication remaining important. Easier creation also raised concerns about validation effort, users' understanding, maintenance, and personalization weakening shared understanding. We discuss seven opportunities for research and practice concerning validation, end-user education, dashboard proliferation and rot, organizational guidance, adaptation, novel visualizations, and accountability for AI-generated content. Our findings connect dashboard evolution with the human and organizational work needed to sustain their use.
\end{abstract}

%% ---------------------------------------------------------------------
%% Teaser
%% ---------------------------------------------------------------------

\begin{figure}
    \centering
    \includegraphics[width=\textwidth]{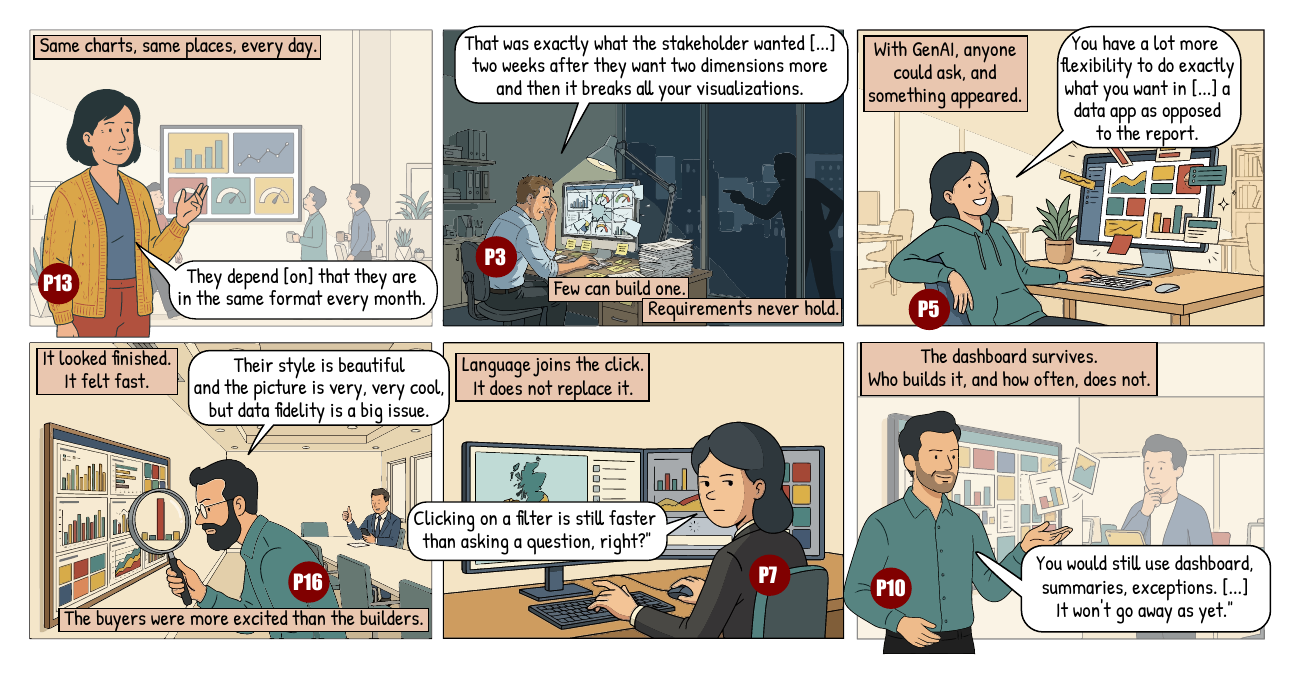}
    \caption{\textbf{The future of dashboards} according to our participants.
    The depicted characters are fictional and bear no resemblance to actual participants.
    All speech bubbles are verbatim participant quotes from our interviews. 
    (Comics by Google Gemini 3.8 Image Flash.)}
  \label{fig:teaser}
\end{figure}


%% ---------------------------------------------------------------------
%% Title
%% ---------------------------------------------------------------------
\maketitle

% \pagestyle{plain}
% \thispagestyle{plain}

%% ---------------------------------------------------------------------
%% Paper Content
%% ---------------------------------------------------------------------

\section{Introduction}
\label{sec:intro}

Visualization dashboards are regularly used to monitor changes, explore data, and support decision-making in various domains~\cite{sarikaya_what_2019, bach_dashboard_2022}.
Behind these dashboards are people who prepare the data, understand what users need, and decide how the information should be presented~\cite{kandel_enterprise_2012, walny_data_2020}. 
Commercial dashboarding systems promised to make this work easier through no-code authoring and self-service analytics.
However, creating a dashboard and enabling people to use it effectively still involve expertise, collaboration, and work across several tools~\cite{tory_data_voice_2023, walchshofer_transitioning_2024}.

With the rise of generative AI (GenAI), the promise of easier dashboard creation has gained renewed attention.
Large language models (LLMs) can generate visualization code from natural-language requests and help users iteratively create visualizations and transform data~\cite{maddigan_chat2vis_2023, wang_data_formulator_2025}.
This also raises questions about whether users need a dashboard at all if they can ask an AI system for an answer.
ThoughtSpot argues that ``dashboards are dead'' and promotes personalized insights delivered where people already work~\cite{thoughtspot_dashboards_dead}, while Tableau argues that dashboards will evolve alongside AI~\cite{tableau_dashboards_evolving_2025}.
This raises questions about the future of dashboards and the role of people working behind them.

From research and practice, we know that it is not enough to build a dashboard; authors work with users to identify meaningful questions, interpret requirements, and connect the analysis to its intended audience~\cite{walny_data_2020, tory_data_voice_2023}.
Users then interact with dashboards to explore the data and develop their understanding~\cite{sarikaya_what_2019, setlur_cooperative_2024}.
Now that dashboards with LLMs integrated have provided the opportunity to phrase questions in natural language, a wide variety of questions arise:
Is a dashboard even the right medium for exploration if questions can be answered directly from the data?
Is dashboard interaction required for exploration?
Are authors required to spend hours eliciting user requirements and translating them into meaningful visual representations?

To investigate these questions, we conducted semi-structured interviews with 16 experts whose work includes dashboard development, consulting, visualization research, and teaching.
We asked about their current challenges with dashboards, their experiences of integrating LLMs into their work, and their expectations for dashboard interfaces and stakeholder roles.
Our overall research questions are:

\begin{itemize}
    \item \textbf{RQ1:} What is the future of dashboards in the era of generative AI?
    \item \textbf{RQ2:} What is the future of the people who create and use dashboards in the era of generative AI?
\end{itemize}

We analyzed the transcripts through thematic analysis with a shared codebook, organizing the results around the two research questions.

For RQ1, almost all participants expected dashboards to persist but change form, and described dashboard interaction itself as part of how people explore and make sense of data.

For RQ2, participants expected authors' responsibilities to shift as more of their work could be delegated to AI, and raised concerns about relying on AI output without understanding it and about maintaining a shared understanding of the data as dashboards become easier to generate and personalize.

This paper contributes an interview study with 16 dashboard experts, spanning academia, industry, and consultancy across 14 countries, on how generative AI is expected to change dashboards and the people who create and use them.
We report on the findings from these interviews and  discuss seven opportunities for research and practice, spanning validation, end-user education, dashboard maintenance, organizational guidance, adaptation, novel visualizations and interactions, and accountability for AI-generated content.
\section{Related Work}
\label{sec:related-work}

To understand the future of dashboards, we look at existing dashboard practices, the role of AI, and the user roles involved.
Therefore, in this section we first discuss dashboard design and organizational practices, then examine how natural language and AI assistance support authoring and interaction.
We then finally explain the user roles and responsibilities involved in AI-assisted analysis using visualizations and dashboards.

\subsection{Dashboard Design, Use, and Organizational Practice}

Dashboards are routinely used in organizations and have also been discussed in the visualization literature.
Stephen Few in his book defined dashboards, highlighted their use cases, and provided design guidelines~\cite{few_information_2006}.
More recently, Sarikaya et al.~\cite{sarikaya_what_2019} discussed what dashboards are, the purposes for which they are built, and the domains they are used in.

Despite this, it is difficult to provide a single working definition of dashboards, as they are used in different ways across different domains.
For example, dashboards can be used for monitoring, exploration, and communication~\cite{sarikaya_what_2019}.
They can also be used for decision-making, where the dashboard is a tool to support the decision-making process~\cite{bach_dashboard_2022}.
For the purpose of our work, we define dashboards as a collection of interactive visualizations on a single screen that are used to communicate data-driven insights and support decision-making~\cite{dhanoa_onboarding_2022}.
We adopt this as a working definition, recognizing that practitioners use the term in different ways.

Beyond definitions, researchers have also studied dashboards as design artifacts.
Bach et al.~\cite{bach_dashboard_2022} discussed dashboard design patterns spanning layout, composition, and interaction, giving authors a shared vocabulary for the structures a dashboard can take.
These patterns matter because a dashboard is read and acted on repeatedly, within the routines of a particular workplace, in the sense of Suchman's situated action~\cite{suchman_plans_1987}.
Elias and Bezerianos~\cite{elias_annotating_2012} found that business-intelligence analysts add and manage annotations on dashboards to record findings and context over time, adding meaning onto a structure that stays recognizable.
Hutchins is his work on distributed cognition~\cite{hutchins_cognition_1995} states that a stable spatial layout offloads memory into the environment rather than requiring it to be held in the head, so the layout itself becomes part of how people think with the dashboard. We discuss the importance of this stability for our findings in Section~\ref{sec:discussion}.

In practice, dashboards are typically designed by experts and handed over to the end users for exploration.
These expert authors are often data science workers who collaborate with the domain users to understand their needs and build dashboards that support their analysis~\cite{kandel_enterprise_2012, kandogan_data_2014}.
Crisan et al.~\cite{crisan_passing_2021} describe data science work as a set of interconnected higher- and lower-order processes spanning preparation, analysis, deployment, and communication.
Commercial dashboarding systems promised to make dashboard creation easier by offering no-code solutions and bringing data science work into a single tool.
This meant that non-experts and domain users would rely less on data and visualization experts~\cite{tory_data_voice_2023, walchshofer_transitioning_2024}.
However, this was not the case, as data practices remained siloed, data science work frequently required switching between multiple tools, and dependence on experts for dashboard creation and use remained~\cite{tory_data_voice_2023, walchshofer_transitioning_2024}.
Tory et al.~\cite{tory_data_voice_2023} found that dashboard users' work involved exploring data, constructing narratives, and communicating with others, while their engagement was often dependent on experts and several tools.

Since the addition of LLMs to the dashboarding ecosystem, these claims have further increased, as LLMs promise to provide dashboards in an instant and automated insights from the data~\cite{thoughtspot_spotterviz}.
Through the use of natural language, users can now ask questions and get answers in the form of visualizations.
We discuss the literature on AI assistance in dashboard and visualization authoring, both using NLIs and exploring other interaction possibilities, and how the roles of humans are changing in the dashboarding ecosystem in the next two sections.

\subsection{AI Assistance in Dashboard and Visualization Authoring} %: NLI and Beyond}

With GenAI, the promise of no-code solutions has been extended to the promise of natural language interfaces (NLIs) that allow users to ask questions and get answers in the form of visualizations~\cite{maddigan_chat2vis_2023, wang_data_formulator_2025}.
Commercial tools such as Tableau, Power BI, and ThoughtSpot now offer natural-language assistance for creating visualizations, report pages, and dashboards~\cite{tableau_agent_documentation, microsoft_copilot_reports, thoughtspot_spotterviz}.
However, the use of AI and natural language in visualization authoring is not new~\cite{wu_ai4vis_2022, narechania_nl4dv_2021}.
NL4DV~\cite{narechania_nl4dv_2021} takes a dataset and a natural-language query, identifies the data attributes and analytical tasks in the request, and returns visualization specifications.
Wang et al.~\cite{wang_towards_2022} extended natural-language authoring to editing operations through a representation that translates requests into executable actions, demonstrated in tools including VisTalk.
With LLMs, Chat2VIS~\cite{maddigan_chat2vis_2023} combines descriptions of the dataset with users' requests to generate Python code for visualizations.
These approaches allow users to describe the analysis or edits they want while the system translates those requests into specifications, actions, or code. 

To understand the intent of the users, Setlur et al.~\cite{setlur_cooperative_2024} drew on conversational processes, including establishing shared understanding and repairing misunderstandings, to develop 39 design heuristics for supporting analytical conversations for dashboards.
Tory and Setlur~\cite{tory_what_2019} studied analytical conversations and emphasized the importance of interpreting intent and context behind a request.
After the intent is clear, authors also need to decide which views belong together and how they should support the intended analysis.
MEDLEY~\cite{pandey_medley_2023} addresses this through recommendations of views and filter widgets based on analytical intentions, such as comparing categories or examining changes. Authors can specify these intentions or select data attributes and views of interest, choose from the recommendations, and configure interactions between the views. BOLT~\cite{srinivasan_bolt_2023} takes this idea to natural-language dashboard authoring by mapping requests to dashboard objectives and recommends content that authors can assemble and customize. 

There is also research on how language can work with other interaction modalities when users interact with visualizations and dashboards.
DataBreeze~\cite{srinivasan_databreeze_2021} combined speech, pen, and touch so that users could move between manually arranging data points and asking the system to organize them.
More recently, InterChat~\cite{chen_interchat_2025} combined natural language with direct manipulation, allowing interactions with visual elements to provide context for analytical requests.
This gives users a way to refer to what they see while describing what they want to do. 
DynaVis~\cite{vaithilingam_dynavis_2024} also connects language with graphical interaction: when users request a visualization edit, the system applies the change and generates a persistent widget that they can use for further adjustments.
Data Formulator 2~\cite{wang_data_formulator_2025} extends this iterative process to data transformations.
Users combine graphical controls and natural language to specify a visualization, while AI generates the transformations needed to produce it.
They can then revisit earlier steps and branch their analysis.
Across these systems, language becomes part of an ongoing interaction in which users can select, adjust, and revisit their choices.

AI assistance also extends to the material or guide that helps users understand a dashboard. 
Plume~\cite{lisnic_plume_2025} supports dashboard authors in composing text with LLM assistance, using the placement of the text and its semantic connection to dashboard content to guide generation.
This helps authors specify what the text should communicate in relation to the visualizations. Hey Dashboard~\cite{dhanoa_hey_2026} approaches this from the user's side by developing DIANA, an assistant for dashboard onboarding.
Users can ask questions through voice or text and point to dashboard elements, while the assistant connects its responses to relevant visualizations through highlighting. 

While the literature supports the use of natural language for lowering the authoring barrier in dashboards~\cite{wang_towards_2022, maddigan_chat2vis_2023, vaithilingam_dynavis_2024, wang_data_formulator_2025}, practitioners and vendors are debating whether dashboards are still needed or will stay the same~\cite{tableau_dashboards_evolving_2025}.
The argument that dashboards are ``dead'' and will soon be replaced by AI-driven insights delivered within users' workflows~\cite{thoughtspot_dashboards_dead, mercurio_dashboards_2026} has surfaced repeatedly over the past few years.
Our work examines this argument and asks what role dashboards will play in the future.

\subsection{Human Roles and Responsibilities in AI-Assisted Analysis}

Traditionally, authors bring together several kinds of expertise to create dashboards.
These authors are not just designing the dashboards, but they are also responsible for preparing the data, connecting the analysis to organizational needs, and ensuring that the dashboards are usable and effective~\cite{walny_data_2020}.
In \emph{Passing the Data Baton}, Crisan et al.~\cite{crisan_passing_2021} described nine data-worker roles with different combinations of expertise in computing, statistics, domain knowledge, and human-centered design.
Their account shows how data work depends on contributions from people with different responsibilities.
The promise of making dashboard creation easier therefore raises questions about how these responsibilities change when more of the work can be delegated to AI.

As AI capabilities expand, the roles of data analysts, visualization designers, and end users are being questioned.
Heer~\cite{heer_agency_2019} examined how automation can support human agency in interactive systems, arguing for designs that let people guide and refine automated assistance.
Dhanoa et al.~\cite{dhanoa_agentic_2025} developed agent-based design patterns from existing visualization systems, providing a framework for considering how work can be distributed between people and agents.
Rogers and Crisan~\cite{rogers_tracing_2023} developed a taxonomy of artifacts produced through human and automated data work and used it to design AutoML Trace.
The visualization makes the sequence and dependencies of human and AutoML contributions visible, demonstrated through a usage scenario with a team developing an AutoML system.
This work highlights the coordination involved when automated processes become part of a larger workflow.

Understanding how these roles are actually changing requires asking the people who hold them, and our approach follows a tradition of interview studies that examine visualization and data practice directly.
Kandel et al.~\cite{kandel_enterprise_2012} interviewed enterprise analysts to characterize the discovery, wrangling, profiling, modeling, and reporting stages of their work.
Alspaugh et al.~\cite{alspaugh_futzing_2019} interviewed 30 professional data analysts about their exploration practices, distinguishing open-ended exploration from exploration as a precursor to directed analysis.
Batch and Elmqvist~\cite{batch_interactive_2018} interviewed data scientists and economists at a government statistics agency to identify where interactive visualization was, and was not, used during initial exploratory analysis.
Newburger and Elmqvist~\cite{newburger_visualization_2024} interviewed statisticians about the role of visualization in their inferential work, drawing on a participant pool with over a hundred combined years of statistical experience.
These studies show that direct engagement with practitioners surfaces expectations, workarounds, and tacit knowledge that documentation and system logs alone do not.
We adopt the same approach to examine how dashboard authors and users expect their own roles to change under generative AI.

Beyond dashboards, there are also empirical studies that examine how data practitioners approach AI assistance in their analysis process.
McNutt et al.~\cite{mcnutt_design_2023} interviewed data scientists to inform the design of AI-powered code assistants for notebooks, examining how assistance could fit into their existing ways of working.
Beschi et al.~\cite{beschi_eud_2025} studied visualization practices through eight interviews in a company and used a prototype to explore expectations of AI-supported end-user development.
Hong and Crisan~\cite{hong_data_chat_2025} examined this interaction during exploratory visual analysis using AI Threads, a conversational interface that supports LLM-generated visualizations and allows users to refine an analysis in separate conversation threads. Across two studies with 50 data workers, they identified three recurring conversational loops: analysis elaboration, refinement, and explanation.

In biomedical visualization, Ziman et al.~\cite{ziman_tensions_2026} interviewed 17 practitioners and researchers about their use and avoidance of GenAI.
They found that visually appealing outputs could contain inaccurate scientific content, making assessment by people with relevant expertise necessary.
Participants' attitudes ranged from enthusiastic adoption to skepticism, and the study highlighted human intervention in both checking scientific accuracy and designing with empathy.
Beyond AI-assisted work, Akbaba et al.~\cite{akbaba_troubling_2023} examined care and the relationships involved in visualization collaboration.
Together, these perspectives draw attention to the responsibilities of understanding the domain users involved and assessing whether a visualization is appropriate for its intended use.

Designing a dashboard is therefore not enough: the responsibilities of authors extend beyond meeting technical requirements.
These interview studies show that data practitioners approach generative AI with a mix of enthusiasm and caution, but they focus on biomedical visualization~\cite{ziman_tensions_2026}, a single organization~\cite{beschi_eud_2025}, or exploratory analysis sessions~\cite{hong_data_chat_2025} rather than on dashboards as durable, shared artifacts.
We complement them with a cross-domain, cross-organizational account of how the dashboard itself and the work of creating and using it are expected to change together.
\section{Method}

We conducted semi-structured interviews with 16 domain experts to understand the current challenges with dashboards, how participants use LLMs in their work, and their expectations for the future of dashboards and stakeholder roles.

\subsection{Participants}
\label{sec:method-participants}

Participants were recruited through professional networks and online communities, all with expertise in dashboard design, development, or use.
Their responsibilities covered building dashboards for clients, developing authoring tools, supporting domain experts, maintaining reporting workflows, and researching and teaching visualization; several worked across these domains.
Participants were based in 14 countries and included 3 female and 13 male participants.
Their experience ranged from about 8 to 25 years in visualization, data, or dashboard work.
Table~\ref{tab:participants} summarizes their sector, role, and demographics.

Our participants observe dashboards from several vantage points, which shape what they notice.
Eight work in academia, researching or teaching visualization; four work in industry, building dashboarding products or maintaining reporting workflows (P2, P5, P8, P13); and four are consultants who deliver or advise on dashboards for clients (P1, P3, P4, P7).
This gives our sample first-hand experience of how dashboarding tools are designed, sold, and adopted, but also perspectives shaped by the commercial platforms and research agendas that our participants are invested in, which we take into account when interpreting their expectations.

% Source: table/Codebook - Demographics.csv.
% Country follows the location recorded in the About column.
% Sector is our classification from each participant's stated role: Academia (universities
% and research groups), Industry (employed at a company on products or research), and
% Consultancy (independent or advisory roles delivering dashboards for clients).
% Sector badge macros (\sectorAca / \sectorInd / \sectorCon) and the row strut (\rowh)
% are defined in main.tex. \rowh keeps every row the same height regardless of content;
% "Role and focus" text is kept short enough to stay on one line.
\begin{table*}[t]
\centering
\caption{Participants' sector, role, and demographic information. F denotes female and M denotes male.}
\label{tab:participants}
\small
\setlength{\tabcolsep}{4pt}
\renewcommand{\arraystretch}{1.15}
\begin{tabular}{@{}p{0.035\linewidth}p{0.10\linewidth}p{0.035\linewidth}p{0.145\linewidth}p{\dimexpr0.585\linewidth-8\tabcolsep\relax}@{}}
\toprule
ID & Country & Gender & Sector & Role and focus \\
\midrule
\rowh P1  & Germany      & F & \sectorCon & Advises on BI strategy and templates; builds client dashboards \\
\rowh P2  & USA          & M & \sectorInd & Industry researcher building natural-language authoring tools \\
\rowh P3  & Denmark      & M & \sectorCon & In-house BI team; helps others build reports \\
\rowh P4  & Brazil       & M & \sectorCon & Builds dashboards as consulting deliverables \\
\rowh P5  & USA          & M & \sectorInd & Software developer building LLM-driven authoring tools \\
\rowh P6  & Japan        & M & \sectorAca & Professor; multidimensional and graph-data visualization \\
\rowh P7  & UK           & M & \sectorCon & Co-founder; writes and speaks on dashboards and AI \\
\rowh P8  & USA          & M & \sectorInd & Industry researcher on language-based dashboard authoring \\
\rowh P9  & Canada       & M & \sectorAca & Professor; visualization and HCI research, prior industry \\
\rowh P10 & India        & F & \sectorAca & Professor; visualization research and teaching \\
\rowh P11 & Austria      & M & \sectorAca & Research group head; visualization for automotive and rail \\
\rowh P12 & Saudi Arabia & M & \sectorAca & Research group head; visual analytics for domain scientists \\
\rowh P13 & Sweden       & M & \sectorInd & Practitioner; client data reporting and maintenance \\
\rowh P14 & Australia    & M & \sectorAca & Professor; network visualization and interactive systems \\
\rowh P15 & South Korea  & M & \sectorAca & Professor; visual analytics systems, including AI agents \\
\rowh P16 & China        & F & \sectorAca & Professor; integrates visualization, machine learning, and AI \\
\bottomrule
\end{tabular}
\end{table*}

\subsection{Interviews}

The interviews were conducted between November 2025 and August 2026.
Each interview lasted approximately 30 to 45 minutes, and sometimes longer, depending on the participant's availability and the depth of discussion.
The interviews were conducted via video conferencing and were recorded and transcribed with the participants' consent.
Before each interview, participants signed a consent and data-protection form that authorized the recording and the use and processing of their data for the study.
We divided the interview questions into three parts.

\paragraph{Introduction and participant background}

In the first part, we introduced ourselves, the purpose of the study, and the terms visualization, dashboard, GenAI, and LLMs.
We explained how the interview transcripts would be used and addressed participants' questions about privacy and the interview.
We then asked participants about their current roles and responsibilities, their pain points with dashboards, and whether they had experimented with integrating LLMs into dashboard experiences.
We also asked about their experiences with hallucinations and their views on natural language interfaces compared with visual elements.

\paragraph{Main interview}

In the second part, we asked how AI had started influencing participants' dashboard and visualization practices, and whether they had observed changes in how people used dashboards or expected them to behave.
We asked which AI features and forms of automation they found promising or concerning.
We also asked how the role of human designers was changing, including whether they expected to continue designing dashboards or to curate and guide AI.
Questions about the future covered how data interfaces might evolve, how people might interact with them through prompts, voice, visuals, gestures, or other modalities, and which parts of the traditional dashboard experience might persist, change, or disappear.

\paragraph{Closing reflections}

In the final part, we asked whether participants expected dashboards to continue to exist, evolve into something else, or give way to other forms of data interfaces.
We asked which human skills and design principles would remain important as AI became more capable, and what advice participants would give to future dashboard designers.
We ended by inviting any additional comments about the future of dashboards or visual analytics.

\subsection{Analysis}

We analyzed the transcripts using thematic analysis with a shared codebook~\cite{braun_using_2006}.
Authors 1 and 2 read eight transcripts closely and coded them, developing descriptive codes and recording notes; Author 1 and Author 3 read and coded the remaining eight transcripts.
The coders then compared their codes, merged those that captured the same idea, and Author 1 then assembled them into a codebook.
We refined this codebook iteratively over several discussions, adding codes for material the existing set did not capture and merging codes that overlapped.
Coding was collaborative rather than independent: we did not compute inter-coder agreement, and we resolved disagreements through discussion.

We then grouped the codes into broader themes, organized around our two research questions: the \emph{future of dashboards}, covering current challenges, LLM integration, and how interfaces might evolve, and the \emph{future of stakeholder roles}, covering the shifting responsibilities of authors and users.
These themes structure Section~\ref{sec:results}.
\section{Results}
\label{sec:results}

We summarize the results of our study of 16 participants around two research questions:
the future of dashboards and the changing roles of the people who create and use them.
For RQ1, we report the current challenges with dashboards, the integration of LLMs, and participants' expectations for the persistence, evolution, and interaction modalities of dashboards.
For RQ2, we report the changing roles of authors and users, the importance of design knowledge and human skills, and the value of human involvement.
We also report participants' concerns about shared understanding, reliance on AI, and cognitive surrender.

\paragraph{Reporting conventions}
We use ``most'' for at least 11 of the 16 participants, ``several'' for five to ten, and ``a few'' for two to four; these counts reflect what participants raised spontaneously in a semi-structured conversation rather than responses to a forced-choice survey.

\subsection{RQ1: The Future of Dashboards}

As our literature review showed, visualization dashboards have been a central part of data analysis for decades.
With the rise of GenAI, their existence remains in question.
To answer this research question, we asked participants about the current challenges with dashboards, how they currently integrate LLMs into their work, and their expectations for the future of dashboards.
We report the current challenges (Sections~\ref{sec:results-current-situation} and~\ref{sec:results-current-llm}) first, as participants grounded their expectations in them, and then turn to the future of dashboards (Section~\ref{sec:results-future}).

\subsubsection{Current Challenges with Dashboards}
\label{sec:results-current-situation}

Visualization dashboards are often created by authors with specialized skills, and then used by people with varying levels of expertise.
In a typical workflow, a dashboard author receives a request from a stakeholder, gathers data, and produces a dashboard that the stakeholder can use to explore the data.
All 16 participants reported using LLMs in some form at the time of their interviews.
Alongside this use of LLMs, they described three recurring challenges in their current practice: (1) understanding the task and audience, and the weak feedback loop between authors and users, (2) the effort of building and maintaining high-quality dashboards, and (3) questioning whether a dashboard is the right medium.

\paragraph{Understanding the task and audience, and a weak feedback loop}

P1, P3, and P15 reported that dashboards are only as useful as the tasks they support, but that identifying those tasks remains challenging.
P1 receives requests to build a dashboard around a dataset without sufficient detail about what users want to see.
P15 described eliciting task requirements and translating them into system requirements, while struggling to obtain time from busy domain experts: \textit{``having correct and appropriate requirements is the most difficult part''.}
The difficulty continues after a dashboard is deployed, because little information flows back to its author.
P1, P5, P7, P9, and P14 reported a lack of feedback from users, which makes it difficult to know whether a dashboard is meeting their needs.
P5 reported that there \textit{``isn't like a strong feedback link between the people that use the dashboards and the people that create the dashboards''.}
P1 and P9 similarly described limited or prohibited usage tracking and situations in which the requester was not the end user.
This highlights the gap between the dashboard author and the users~\cite{dhanoa_onboarding_2022}, which can lead to dashboards that do not meet the needs of the intended users.

\paragraph{The effort of building and maintaining high-quality dashboards}

P4, P6, and P10 reported that creating a high-quality dashboard is a time-consuming process, especially when the data is complex or the requirements are unclear.
P4 said that, for them, conceiving a nice dashboard graphically is hard and it is difficult to choose the type of plots from the \textit{``pool of options''}.
P2 and P12 added that the effort depends on the domain, with scientific and multi-layered data requiring more time than business data.
P1, P3, P4, P5, P13, and P16 reported that maintaining and updating dashboards is an equally demanding challenge, especially when the data or requirements change.
P3 described how additional requests disrupted an initially satisfactory dashboard:

\begin{quotebox}[P3]
    ``That was exactly what the stakeholder wanted, of course, two weeks after they want two dimensions more and then it breaks all your visualization and just waste your time of rebuilding that.''
\end{quotebox}

P4 said that traditional dashboard structures do not support data processing well, so they seek alternatives outside dashboard tools after encountering processing limitations.
P13 described the fragility of dashboards, which break entirely when even one dependent data source fails.
P1 and P5 also described the ongoing cost of keeping up with platforms that release updates every month and carry years of backward compatibility.
These challenges highlight the need for better support in dashboard structures to ease the building and maintenance process.

\paragraph{Questioning whether a dashboard is the right medium}

P7, P8, P9, and P14 reported that dashboards are not always the best medium for communicating information.
P8 said that people are trying to answer a question or accomplish a task, rather than wanting a dashboard as such.
P7 said that even after 10--15 years of self-service tools, authors still struggle to build dashboards that answer users' questions.
P9 described dashboards as a poor vehicle for synchronous conversations, and said that bringing one to a meeting made the audience feel as though they were watching a tool demonstration rather than hearing findings or a story:

\begin{quotebox}[P9]
    ``Dashboards are a poor vehicle for synchronous conversations and presentations... If people brought a dashboard to a meeting that people then suddenly felt that they were watching a tool demo and not being told a story or presented findings, so it was just an inappropriate for that medium for that type of that scenario, that communicative scenario.''
\end{quotebox}

Despite these challenges, participants reported that they continue to author dashboards.
They framed these challenges as the baseline that GenAI would need to improve upon.

\subsubsection{Current Integration of LLMs in Dashboards}
\label{sec:results-current-llm}

We next examine where participants used LLMs within their dashboard work and how they experienced that assistance.
They discussed: (1) how they use LLMs in their authoring process, (2) the illusion of working fast, (3) polished but hard-to-verify visual output, and (4) the difference in attitude between them and other user roles.

\paragraph{Use of LLMs in the authoring process}

Most participants (P1--P7, P11--P14, and P16) reported using LLMs in their authoring process to assist with tasks such as generating code, creating mock-ups, and providing design suggestions.
P1 used LLMs to produce early prototypes after discussing requirements with a customer, to assist with coding, and to critique dashboard screenshots.
P14 reported using AI-generated mock-ups in place of manual layout work.
To reduce hallucinations, P2 and P5 constrained what the model was allowed to do, for example by restricting it to a single data source, although P5 noted that instructing a model \textit{``not to hallucinate''} is \textit{``not the end all be all''}.
Adoption was nevertheless uneven: P12 remained doubtful that AI-generated dashboards in their current state are good enough to be used as-is, and P13 reported that their organization has not moved toward AI because its workflow is rigid and repetitive and its client data is confidential.

\paragraph{The illusion of working fast}

Participants described an illusion of working fast when using LLMs in their dashboard work.
P4, P5, P7, P8, and P11 reported that LLMs improved the quality of their output or let them attempt work they would not otherwise take on, without necessarily reducing the time the work took.
P4 described improved quality without reduced processing time:

\begin{quotebox}[P4]
    ``If I spend less time doing data processing, the answer is no. But if I can say if you have the same time that I used before, I get much better quality now.''
\end{quotebox}

P5 described being able to attempt work they would not normally do, without necessarily reaching the result faster.
P8 reported time savings but uncertainty about the predictability of output quality.
P7 questioned whether feeling productive during several hours of prompting meant that the work had actually taken less time.
Participants thus distinguished gains in capability or quality from reductions in effort.

\paragraph{Polished but hard-to-verify visual output}

P1, P5, P6, P7, P8, P10, and P16 reported that AI-generated visualizations look convincing while being difficult to verify.
P16 described difficulty using LLMs for chart generation because attractive outputs did not reliably preserve the data:

\begin{quotebox}[P16]
    ``Their style is beautiful and the picture is very, very cool, but that data fidelity is a big issue.''
\end{quotebox}

P7 recalled a generated dashboard mock-up that contained an incorrectly hard-coded KPI, and another chart that looked correct but had sorted the months alphabetically behind a numeric-looking axis.
P6 and P8 expressed caution because hallucinations are difficult to identify in a visualization, where the error is not visible on the surface, echoing the challenges described by Ziman et al.~\cite{ziman_tensions_2026}.
P14 worried that an answer could appear plausible while the sources and analytical process behind it remained opaque.
P2 added that natural-language phrasing raises expectations, as some users assume a correct answer simply because they could formulate the question.
We return to what this means for the author's role in Section~\ref{sec:results-roles}.

\paragraph{Difference in attitude between them and other user roles}

Participants also described differences between those who select tools and those who work with them.
P3, P5, and P9 reported that dashboard creators are skeptical about the benefits of LLMs, while the people responsible for buying and rolling out the tools are enthusiastic about the anticipated time savings.
P9 reported that the excitement of executives was not always shared by analysts encountering limitations in everyday use.
P3 noted that the enthusiasm often came from influencers rather than the users or report builders.
Prior experience with a dashboarding tool also shaped how participants received AI-generated artifacts.
P5 pointed out the difference between experienced users, who needed to work out how an AI-generated artifact operated, and newer users, who could appreciate the flexibility of a generated data application:

\begin{quotebox}[P5]
    ``I feel like there's two buckets of people. There are people that are used to [the platform] and people that are not used to [the platform]. ... if something's built with AI, those power users... they have to try to figure out what's going on. For the users that are newer to [the platform], it's almost swapped. Because you have a lot more flexibility to do exactly what you want in this, like what we're calling a data app as opposed to the report.''
\end{quotebox}

\subsubsection{Expectations for the Future}
\label{sec:results-future}

Almost all participants expected dashboards to persist but to evolve in their design and functionality, with the specifics depending on the tasks a dashboard supported, established organizational practices, and users' expertise.
We report what participants told us about (1) the persistence of dashboards, (2) their evolution, (3) future interaction modalities, (4) dashboard interaction as an inherent human process, (5) future concerns, and (6) the loss of shared truth through dashboard proliferation.

\paragraph{Persistence of dashboards}

Most participants (all except P6, P11, and P15) expected dashboards to persist, for two kinds of reasons.
The first concerned the task.
P7, P10, and P13 expected dashboards to remain in use for recurring questions, monitoring, and regular reporting.
P7 said that there are questions about sales or leads that users want to check every day or week without asking an LLM each time.
P10 gave the example of watching the World Cup, where viewers would want to see the scores and plots directly during a game rather than repeatedly asking an LLM for details.
P13 expected dashboard creation to become faster while the end product stayed the same, because clients depend on a consistent format every month:

\begin{quotebox}[P13]
    ``And they depend that they are in the same format every month. So I do not think end product changes''
\end{quotebox}

The second reason concerned the organization.
P1, P3, and P9 expected existing dashboards to persist because of their installed customer base and the habits of the organizations using them, even though P9 was skeptical about new dashboard-centric products.

\paragraph{Evolution of dashboards}

Several participants (P9, P12, P14, P15) described how the form of dashboards could change.
P12 expected AI to let users change or generate visualizations according to their needs, rather than being limited to the views chosen by the author.
P15 similarly expected dashboards to become more dynamic, with parts of the dashboard changing when users ask to see something different.
P9 described smaller applications created for a particular question or conversation, and hybrid dashboards that surface answers through annotations or temporary views.
P14 expected dashboards to become more ubiquitous through situated displays that embed information in the surrounding environment, and anticipated that wearable displays would eventually create new challenges of where and when to place a dashboard, ``the same way that phones changed everything''.

\paragraph{Expansion of interaction modalities}

Participants expected more ways of interacting with dashboards, including text, speech, and gestures, while graphical interfaces would remain central.
P2 and P5 expected speech to become more common, although P5 still expected most dashboard use to involve viewing the information.
P15 discussed combining spoken language with gestures, and P6 and P14 saw a role for gestures in 3D visualization, where pointing to a direction or a surface can be easier than describing it in words:

\begin{quotebox}[P14]
    ``I think ideally natural language definitely has its place, but I think there's situations where language alone is a very awkward interface. And especially in spatial kind of data where you need to, where, yeah, gestures are a much more efficient way to indicate a quantity or something, or a direction, or a surface within a volume or something like that''
\end{quotebox}

P5 also expected textual summaries alongside visualizations~\cite{sultanum_instruction_2025}, and P10 raised the possibility of multilingual dashboards:

\begin{quotebox}[P5]
    ``I do foresee text having a heavier role... having like a textual description alongside the visual representation seems to be pretty big because before you only basically had the visual representation and you kind of depended on the person looking at it to understand it.''
\end{quotebox}

For the graphical interface itself, P11 and P12 expected the appropriate complexity to depend on the user: simpler for general users, and richer for professionals or those asking detailed questions through an LLM.
P11 added that using an LLM for small adjustments would still depend on having good visualization defaults, without which the views had to be configured manually.

\paragraph{Familiar controls can be quicker than a prompt}

Participants did not expect these modalities to replace existing interactions.
P7 said that ``clicking on a filter is still faster than asking a question, right?'', giving the example of selecting Scotland with one click instead of typing a request and waiting for a response.
P12 expected mouse and touch interaction to remain common for everyday tasks, with other modalities reserved for more specialized work.
P10 also expected other modalities to remain niche for the general population:

\begin{quotebox}[P10]
    ``I do think even though multi-modality is there, that's somewhat very niche. But if I look at the general population, you would still use dashboard, summaries, exception. That's what I believe. And I think its history of time has shown that that will stay. It won't go away as yet.''
\end{quotebox}

P8 and P9 discussed AI working with direct manipulation, where actions in the interface could signal what the user wanted without a separate prompt.
Participants thus saw the choice of modality as depending on the task, the effort involved, and the control that users needed, giving language and graphical controls complementary roles: users can describe a change and then refine it through controls that provide immediate visual feedback~\cite{vaithilingam_dynavis_2024}.

\paragraph{Dashboard interaction as an inherent human process}

P5 and P8 expected dashboard interaction to remain important because users still need to explore and understand the data themselves.
P5 said that cross-highlighting, cross-filtering, and slicers would continue to be used, and described the interaction itself as a human process:

\begin{quotebox}[P5]
    ``You're not gonna have the LLM step through the interaction or something like that, because that is fundamentally like a human process.''
\end{quotebox}

P8 similarly said that without this interactivity there would be little to distinguish a dashboard from a traditional single-view interface.
This did not exclude AI from exploration: P5 saw an LLM as a collaborator that could suggest questions or directions while the user remained the one making sense of the data, and described recognizing what one does not yet understand as a distinctly human part of the process.

\paragraph{Future concerns}

Participants raised concerns about what would happen as AI took on more of the analysis.
P14 worried that reducing human involvement could make it harder to notice gaps between an AI system's output and its effects on people, and expected users to need explanations of how an answer had been reached.
P3 and P15 raised security concerns: P3 recounted an account of giving an AI system extensive permissions on a personal machine, and P15 imagined agents in critical visual analytics systems becoming targets for attacks.
These were anticipated risks rather than incidents participants had observed in their own dashboard work.

\paragraph{Loss of shared truth and dashboard proliferation}

Several participants (P2, P5, P12, P15) discussed the consequences of making dashboards easier to create and personalize.
P15 expected lower implementation costs to make personal visual analytics systems more common, and P2 described a possible ``death of a thousand dashboards'', where so many are generated that nobody can use them.
P5 recounted how easier report creation had already produced many reports showing the same information, followed by efforts to certify which ones were authoritative.
P2 and P7 additionally worried about a loss of shared truth: when people look at the same data through their own personalized dashboards, the representations, and potentially the conclusions, can differ.
P7 noted that a personal dashboard ``for an audience of one'' is fine until it is moved into production.
This raises the question of how people maintain a shared understanding of the data when each has their own dashboard.

\subsection{RQ2: The Future of Stakeholder Roles}
\label{sec:results-roles}

We also asked participants about the future of stakeholder roles in the dashboard ecosystem.
Participants discussed the future role of analysts and authors, the knowledge they would still need, the value of human involvement, and the changing role of end users.

\paragraph{Shift from direct design to organizing and curating the work of AI}

Most participants (P1, P3, P4, P5, P7, P10--P12, P14--P16) expected authors and analysts to spend less time on implementation as AI took on more of it, and more time directing and reviewing that work.
P4 expected their role to focus on setting requirements, testing the output, and understanding what agents could do.
P7 described moving from producing dashboards to curating the work of agents, provided the technology succeeded:

\begin{quotebox}[P7]
    ``So I think the analyst role is going to change from the producer to the curator.''
\end{quotebox}

P12 already used AI to generate code rather than relying on developers for some of this work, but still expected designers to decide which outputs were suitable, and P5 expected information design to remain important even as the workflow changed.

\paragraph{Shift from technical work to meaningful questions}

As implementation became cheaper, several participants (P2, P5, P7, P11--P13, P15) expected the author's attention to move to why something is being visualized and for whom.
P7 said analysts would need to help LLMs understand the context of users' questions and to recognize when that context had been misunderstood.
P2 described the importance of understanding a person's intent and knowing which analytical approaches were possible, and P5 pointed out that recognizing a gap in one's own knowledge was a human responsibility.
P9 said that building a dashboard would become easier, but understanding the questions and learning goals of its users would still require direct communication:

\begin{quotebox}[P9]
    ``And you're not going to be able to get that solely through prompting. You need to talk to people.''
\end{quotebox}

\paragraph{Knowledge of basic design principles}

Several participants (P4, P5, P7, P10--P13, P16) expected authors to keep needing knowledge of design and communication even when AI generated the visualizations.
P10 said design principles would remain relevant because visualizations are still consumed by humans, and P5 emphasized understanding the audience, the data, and the task, and how to communicate the intended message:

\begin{quotebox}[P5]
    ``Design and understanding of the audience, all the stuff that goes into creating a good visualization, it could even be like abstraction of the data characteristics and the analysis task. Like all this stuff, I don't think any of it really goes away.''
\end{quotebox}

P16 added that designers would need to understand how AI models work in order to use them well.
P11, however, was uncertain how long humans would retain an advantage in generating ideas:

\begin{quotebox}[P11]
    ``I think humans are better in this visionary things...But then when it comes to execution, it outperforms human certainly already now. But you have to have a good idea. And I think these good ideas are still where we are better. But how long I have, I can't say. How long it will be until it generates you need no ideas?''
\end{quotebox}

\paragraph{Importance of experience, empathy, and creativity}

Beyond technical skills, several participants (P1--P4, P7, P8, P10, P11, P14) pointed to the judgment authors bring to dashboard work.
P1 described critical thinking as connecting what people said with experience from other contexts, and expected this to remain important as AI became part of the work.
Understanding the people asking the questions also involves listening and empathy; without these, P4 said, we end up with beautiful dashboards that no one understands or cares about.
P7 highlighted the same qualities alongside creativity:

\begin{quotebox}[P7]
    ``Oh, human empathy, listening skills, critical skills, creativity. Yeah, LLMs, they look like they're amazing, right? It's incredible, but they're just regurgitating other stuff.''
\end{quotebox}

P3 emphasized that the right answers and solutions can involve feelings as well as arguments:

\begin{quotebox}[P3]
    ``Maybe something about empathy. Often, the right answers and solutions are not about arguments, it's about feelings.''
\end{quotebox}

Participants also emphasized developing their own ideas.
P7 questioned the originality of LLM output, and P14 worried that if people stopped sketching and exploring ideas independently, the result could be greater uniformity and a loss of innovation.

\paragraph{Value of human involvement, and its cost}

Participants expected humans to stay involved in evaluating and approving dashboards.
P8 stressed the importance of critically evaluating AI output as it became easier to generate at volume, and P7 pointed out that someone still has to discuss a dashboard with stakeholders and sign off before it goes into production:

\begin{quotebox}[P7]
    ``Physical things is just managing the process, right? Somebody has to sign off on a production dashboard... I have to go to the stakeholder and we have to have a conversation. Do you approve this and should we move to production? So there's still administrative tasks and there's still creative tasks. And you just can't rely on the AIs and the LLMs yet to be accurate.''
\end{quotebox}

Keeping humans in the loop also creates work.
Several participants (P3, P4, P6, P7, P10, P14) noted that generating many candidate designs makes evaluating them laborious, so the author's effort shifts toward checking output and deciding whether it is ready to use.

\paragraph{Future role of end users}

Several participants (P1, P2, P4, P5, P13) expected end users to gain more opportunities to explore data without contacting an analyst.
P4 described how simple prompts could make exploration easier, and P13 expected users to ask basic questions through a chat interface instead of emailing the analytics team.
The gap between authors and users could therefore narrow for some tasks.
However, P7 said that the people best able to use natural-language data interactions were already good data analysts, and P4 questioned whether users who struggle to explore a dashboard could ask an LLM the right questions.
Differences in analytical knowledge would thus remain important even as the tooling gap closed.

\paragraph{Challenges of cognitive surrender}

Several participants (P2, P4, P7, P8, P10, P14) worried about users and authors relying on AI without understanding the work behind its output.
P10 reported that students built more complete applications than in previous years but struggled to explain them.
P2 said that asking directly for an answer could mean thinking less about the analytical steps needed to reach it, and P4 worried that people would stop learning the basics as more of the work was done by AI.
P8 added that as authors offload more to AI, less thought goes into the dashboards that are produced.
The shared concern was that the ability to produce a dashboard would not be matched by an understanding of how it worked.
\section{Discussion}
\label{sec:discussion}

\paragraph{Dashboards persist but evolve}
RQ1 asked about the future of dashboards.
Across our interviews, most participants said that dashboards will persist but evolve.
Widely used dashboards, such as public-facing and monitoring dashboards, would remain largely as they are (P1, P3, P10), while analytical dashboards would change the most.
P9 was the most skeptical, doubting that new dashboard-centric products would emerge while still expecting existing dashboards to persist because of their customer base and the habits of the organizations using them.
Participants expected the change to move toward adaptive dashboards that adjust to user needs (P12, P15) and support a wider range of interaction techniques, including gestures (P6, P14, P15).
Such adaptive and generated forms strain the common definition of a dashboard as a fixed collection of views on a single screen.
The claim that dashboards are ``dead''~\cite{thoughtspot_dashboards_dead, tableau_dashboards_evolving_2025} therefore does not match what our participants described.
As P7 put it, \textit{``dashboards are dead. Along comes another term describing exactly the same thing''}: what changes is the method of production, not the underlying need for a dashboard.

\paragraph{Easier creation redefines user roles and responsibilities}
RQ2 asked about the future of the people who create and use dashboards.
Participants expected generative AI to make dashboards easier to produce, but they did not expect the work around a dashboard to shrink, only to change.
Most participants anticipated a shift from producing dashboards to curating them: less time spent on implementation, and more on specifying requirements, choosing among generated options, checking them, and taking responsibility for what goes into production.
This redistributes effort toward the parts of the process that depend on human judgment and on talking to people, which participants expected to remain difficult to delegate.
Their accounts of working with LLMs also qualified the promise of faster creation: several described better output or the confidence to attempt more, but not less time spent, an \emph{illusion of working fast} that separates gains in capability from reductions in effort.
Participants also emphasized that understanding the audience, critically evaluating AI output, and knowing basic design principles would remain essential to dashboard work (P5, P8, P10).
Even so, they expected the roles of the analyst and the dashboard author to keep changing, as roles have in other domains.

\paragraph{The Stability-versus-Specificity Argument}
The purpose of a persistent dashboard is a familiar, memorable layout.
Daily or more frequent use builds a work practice around fixed gauge locations, in Suchman's sense of situated action~\cite{suchman_plans_1987}.
Hutchins's distributed cognition~\cite{hutchins_cognition_1995} makes the stronger version of the claim: the spatial layout offloads memory into the environment, so regenerating the dashboard destroys an external cognitive scaffold rather than merely disrupting a habit.
A dashboard that is new every time forces re-onboarding on every view.

The resolution splits by task.
Monitoring keeps the stable, familiar dashboard; that use case does not go away.
The one-size-fits-all dashboard pressed into service for every analytical question can go, and good riddance, because a view built for the question at hand is a real gain.
For just-in-time dashboards, the recommendation is stable structure with generated content: the generative system fills a persistent scaffold instead of inventing a fresh layout each time.
Bach et al.'s dashboard design patterns~\cite{bach_dashboard_2022}, consistency principles from coordinated multiple views, and layout-template work in visualization grammars all give this middle path precedent.

\section{Opportunities for Research and Practice}
\label{sec:opportunities}

With GenAI, making dashboards has become easier, but it also raises questions about what happens after they are generated.
Someone still needs to check whether the dashboard is correct, understand how to use it, and maintain it as the data and requirements change.
Now that users can create their own dashboards using natural language, the management responsibilities may also extend to these domain users who previously relied on an analyst or an expert.
Based on the participants' observations and expectations, we discuss seven opportunities for research and practice.
These opportunities concern validation, end-user education, dashboard maintenance, organizational guidance, adaptation, the place of novel visualizations and interactions, and accountability for what AI generates.
We propose them as directions to investigate, grounded in the challenges and possibilities raised in the interviews, and summarize them in Table~\ref{tab:opportunities}.

\begin{table*}[t]
\centering
\caption{The seven opportunities, the observations that motivate them, and the directions they suggest for research and for practice.}
\label{tab:opportunities}
\footnotesize
\setlength{\tabcolsep}{4pt}
\renewcommand{\arraystretch}{1.15}
\begin{tabular}{@{}p{0.15\linewidth}p{0.25\linewidth}p{0.27\linewidth}p{0.24\linewidth}@{}}
\toprule
Opportunity & What participants observed & Direction for research & Implication for practice \\
\midrule
Making human validation manageable (\S\ref{sec:opportunity-validation}) & Generated output can exceed what a person can check, and subtle errors still look correct (P5, P7, P10, P14). & Whether in-dashboard provenance and change summaries help people locate errors and target their review. & Combine automated checks with interfaces for reviewing edits and recording approval. \\
\addlinespace
Helping end users recognize mistakes (\S\ref{sec:opportunity-literacy}) & Users can produce dashboards they cannot explain or judge (P4, P6, P10). & Whether learning offered while people work with their own data improves later, unaided error detection. & Pair creation features with scaffolded, in-context learning about encodings, scales, and filters. \\
\addlinespace
Managing proliferation and rot (\S\ref{sec:opportunity-maintenance}) & Easier creation multiplies near-duplicate dashboards whose data and assumptions go stale (P1, P2, P5, P12, P15). & Longitudinal study of which generated dashboards are reused, who owns them, and how rot is noticed. & Make ownership, data dependencies, and lifecycle status visible; support consolidation and retirement. \\
\addlinespace
Developing organizational guidance (\S\ref{sec:opportunity-governance}) & Data sensitivity and client expectations limit where AI can be used (P1, P13). & How organizational policy and regulation can be translated into guidance authors and users can act on. & Team-level rules on which data may be processed, where, and who authorizes it. \\
\addlinespace
Balancing adaptation with shared understanding (\S\ref{sec:opportunity-adaptation}) & Adaptive and personalized views help individuals but can erode a shared reference (P2, P9, P12, P15). & Which elements should stay stable, and whether people can reconstruct the views behind a conclusion. & Keep a shared reference view; make personal adaptations visible and reversible; share exact selections. \\
\addlinespace
The future of novel visualizations and interactions (\S\ref{sec:opportunity-novel-visualizations}) & Advanced techniques are admired in demonstrations but rarely adopted in practice (P6, P11, P14). & Long-term study of whether and why techniques are retained, and what support they need. & Evaluate novelty, complexity, and unfamiliarity separately when deciding what to offer. \\
\addlinespace
Accuracy, hallucination, and accountability (\S\ref{sec:opportunity-accountability}) & Polished output hides errors, and the consequences fall on people downstream (P2, P7, P10, P16). & Accuracy expectations by use; how to disclose AI-generated content and its provenance; how sign-off is distributed. & Mark AI-generated views, retain provenance into production, and define who approves and is answerable. \\
\bottomrule
\end{tabular}
\end{table*}

\subsection{Making Human Validation Manageable}
\label{sec:opportunity-validation}

Keeping a human in the loop becomes difficult when the amount of generated work exceeds what a person can reasonably evaluate.
It is especially challenging when the visualization seems correct but contains subtle errors~\cite{ziman_tensions_2026}. 
Without the ability to check the underlying data, calculations, and visual encodings, it is difficult to know whether a dashboard is correct.
There are already attempts to share this work with AI: P5 described using an LLM as a judge alongside weekly human testing sessions.
However, this leaves a question about how much checking people still need to do and what would help them do it.
For dashboards, validation can involve checking individual values, their calculations, the visual encodings, and whether the views together answer the intended question.

An option could be to make the information needed for these checks available within the dashboard.
P14 wanted visualizations to explain where an AI-generated answer came from and allow people to interrogate it.
This could take the form of connecting a displayed value to its source data, filters, and calculations, or showing which parts of a dashboard changed after an AI edit.
Future research could investigate whether these forms of support help people find errors and decide which parts require closer inspection.
Additionally, commercial dashboarding tools could combine automated checks with interfaces for reviewing changes and recording human approval.
Such support would need to be evaluated in terms of the effort involved and the errors that remain undetected despite these measures. 

\subsection{Helping End Users Recognize Mistakes}
\label{sec:opportunity-literacy}

If end users can generate their own dashboards, what knowledge will they need to judge what they have created?
P10's experience of students producing more complete applications while struggling to explain them shows why the ability to create an output does not establish an understanding of it.
P4 raised concerns about the limited place of visualization in school education, and P6 emphasized the need to learn how to judge the validity of data through visualizations.
This creates an opportunity to consider end-user education as part of dashboard authoring.
Users would need support in understanding how choices about scales, aggregations, filters, and visual representations affect what a dashboard communicates.

This education could also happen while people work with their own data.
P4 suggested that an assistant could guide users through graphs and tabs, while P5 asked, ``I wonder if there's an opportunity to help prime people for [showing] how you're intended to use a dashboard.''
Existing dashboard onboarding work supports users in understanding dashboard content and interactions~\cite{dhanoa_onboarding_2022, dhanoa_hey_2026}.
As users take on authoring tasks, this support could extend to examining how a generated dashboard was constructed.
For example, an authoring interface could help users compare a displayed total with the underlying records or examine how a filter changes the population represented.
Future research could examine whether these activities improve users' ability to identify mistakes in subsequent dashboards without assistance.
For industry, this means considering learning alongside the features that make creation easier, so that users have opportunities to develop the judgment needed to assess their own work.
This could also potentially lower the burden on authors, who would otherwise need to check every dashboard created by end users.

\subsection{From Heirloom to Fishwrap: Managing Dashboard Proliferation and Rot}
\label{sec:opportunity-maintenance}

Several participants (P2, P5, P12, P15) foresee that easier dashboard creation will lead to dashboard proliferation.
P2 worried about generating more dashboards than people could use.
P5 explained that they had already started seeing this trend: easier report creation led to duplicate reports and was followed by efforts to certify them.
Maintaining these reports also means preserving the work already built on a platform.
P5 described having to support years of backward compatibility, including retaining design decisions because changing them could disrupt existing reports.
P1 similarly described the effort of keeping up with frequent platform updates.
These experiences suggest that the cost of creating a dashboard is only one part of the work required to keep it useful.

As generation becomes easier, there is an opportunity to investigate whether dashboard proliferation increases this maintenance burden and contributes to dashboard rot, where dashboards remain available while their data, dependencies, or assumptions become outdated. 
Longitudinal research could follow generated dashboards to examine which are reused, who takes responsibility for them, and how errors or obsolete requirements are discovered.
Industry could support this work by making dashboard ownership, data dependencies, and maintenance status visible, and by helping authors identify reports that could be consolidated or retired.
An option could be to implement a dashboard lifecycle management system that automatically flags and archives outdated dashboards.
For instance, temporary dashboards created for a single question will have a different lifecycle from reports used repeatedly across an organization.
Authoring tools could therefore help users decide whether a generated dashboard is a temporary exploration or a maintained data product, and what responsibilities follow from that choice.

\subsection{Developing Organizational Guidance for AI-Assisted Dashboards}
\label{sec:opportunity-governance}

The opportunity to use AI in dashboard creation also depends on what organizations are willing and permitted to do with their data.
P13 worked with confidential client data and explained that clients were pleased that the organization was not moving heavily toward AI because they did not want their data exposed to outsiders.
P1 similarly noted that the companies they work with were often reluctant to adopt AI tools with their data.
For these workflows, deciding whether to use AI involved responsibilities to the people supplying the data.
As AI becomes part of authoring and interaction, dashboard teams need guidance about which data can be processed, where processing can take place, and who can authorize it.
Future research could focus on long-term exploration with dashboard teams, data owners, and privacy specialists to investigate how organizational policies and applicable regulatory requirements can be translated into guidance that authors and users can act on. 

\subsection{Balancing Adaptation with Shared Understanding}
\label{sec:opportunity-adaptation}

As participants P12 and P15 anticipated (Section~\ref{sec:results-future}), dashboards may increasingly adapt to individual users, generating or reshaping views on demand. On the one hand, it fixes the rigidity that comes with the dashboard by allowing it to accommodate users whose analytical goals shift over time. But on the other hand, it can lead to a loss of shared understanding. P2 raised this as a concern about what happens when everyone looks at the same data through a different personalized dashboard.
Adaptation therefore creates a tension between supporting individual needs and maintaining a shared reference for discussion, also discussed in Section~\ref{sec:discussion}.
For adaptive dashboards, different representations do not necessarily mean conflicting conclusions, but it needs to be clear to the users what is different and how can it affect their interpretation of the data.

P9 suggested that annotations and temporary secondary views could be one of the ways to accommodate new questions while retaining a familiar dashboard.
This opens up questions about which elements should remain stable, which can adapt, and when users should control a change.
Changes to layout, displayed measures, and interaction controls may have different effects on users' ability to orient themselves and work together.
Researchers could examine these effects in recurring use and collaborative tasks, including whether people can reconstruct and compare the views behind a conclusion.
In practice, dashboard tools could retain a shared reference view while making personal adaptations visible and reversible, similar to how version control systems work.
They could also allow users to share the exact data selection and view used in a discussion.
The challenge is to establish how much adaptation is useful for a particular task while keeping the basis of the analysis understandable to others.

\subsection{Considering the Future of Novel Visualizations and Interactions in Dashboards}
\label{sec:opportunity-novel-visualizations}

We, as visualization researchers, often balance supporting users' tasks with creating novel visualization and interaction techniques.
In the short term, these techniques may be well received during evaluations and demonstrations, but we know less about whether people continue to use them in practice.
P14 described this challenge by learning that users' needs were sometimes simpler than anticipated: \textit{``not everything needs to be a completely novel visualization or a different way of interacting with the data.''}
P11 described offering advanced interactions, such as combining brushes through Boolean operations, that users appreciated when demonstrated but used infrequently in practice.
This raises a question for our research field: what place will novel visualizations and interactions have when familiar techniques already meet many users' needs?

If AI reduces the effort of implementation, it could theoretically become less challenging to make different techniques available within dashboards.
However, making a technique available does not mean that people will understand it or find it useful.
P11 described trying to educate users about advanced interactions, while also questioning whether enough people needed them.
Future research could therefore examine whether limited use reflects a need for explanation and support, or whether a (novel) technique offers little benefit for the users' tasks.

What would this mean for how we evaluate novel visualizations and interactions?
Long-term evaluations could examine whether people return to a technique after its initial demonstration, which tasks lead them to use it, and why they stop using it.
Such studies could also investigate the assistance people need over time.
For example, do users continue to use advanced brushing or combinations of language and graphical controls when familiar controls are also available?
Following these choices in everyday dashboard work could help us understand when a technique becomes useful beyond an initial demonstration.
Novelty, complexity, and unfamiliarity would need to be considered separately, since a novel technique is not necessarily complex, and an established technique may still be unfamiliar to a user.

\subsection{Accuracy, Hallucination, and Accountability}
\label{sec:opportunity-accountability}

In order to support human validation on AI output, it is important to understand what needs to be checked and how. What counts as \textit{accurate} enough? How \textit{errors} are disclosed? And who is \textit{answerable} when a generated dashboard informs a decision and later turns out to be wrong?
P7 talked about a mock-up with a hard-coded and incorrect KPI, and P16 said that visually appealing output from current models still has a \textit{``data fidelity''} problem.
Because a dashboard presents its output as authoritative, these errors are harder to notice than in prose, and P2 observed that a natural-language interface can lead users to assume a correct answer simply because they were able to ask the question.
The consequences fall on people downstream: P10 noted that policy-makers may act on dashboards and textual summaries that state what the data does not show.

Research could establish what accuracy expectations are appropriate for different dashboard uses, since a glanceable indicator, an exploratory view, and a public report all carry different risks.
It could also examine how to communicate that a view or summary is AI-generated, and with what confidence, so that users neither over-trust nor dismiss it, and whether provenance and change history can be carried from authoring into deployment, extending work that makes human and automated contributions visible~\cite{rogers_tracing_2023}.
It is also worth questioning how responsibility should be distributed among the author who curated the output, the organization that deployed it, and the end user who generated it, particularly as the last of these becomes more common.
In practice, dashboarding tools could mark AI-generated views and summaries, retain their provenance into production, and define approval roles so that, as P7 put it, someone still signs off on a production dashboard.
These measures would need to be evaluated for whether they change how much users trust a dashboard and how reliably errors are caught before they matter.

We offer these as directions grounded in what experts currently expect rather than as validated solutions, and each requires empirical work to establish whether the proposed support helps and at what cost.
\section{Limitations}
\label{sec:limitations-future-work}

\paragraph{Sample representativeness.} 

In this work, we bring together the perspectives of 16 experts whose responsibilities span dashboard development, consulting, research, and teaching. All participants reported using LLMs in some form. The study therefore offers limited insight into the perspectives of people who have chosen not to use these tools. Although participants were based in 14 countries, this geographical spread does not make the sample representative of dashboard practice across those countries.

\paragraph{Perspectives of other stakeholders.} 

We focused on people with expertise in creating, researching, or supporting dashboards. Their experiences with clients, students, and colleagues informed their observations about other stakeholders, but these observations cannot substitute for hearing directly from those stakeholders. In particular, our findings about end users' needs, managers' expectations, and organizational decisions largely reflect the experts' perspectives. Participants also drew on different settings, including business reporting, scientific visualization, and visual analytics. We have taken this into account while reporting their examples, as an expectation concerning a specialized analytical task may not apply to everyday dashboard use.
Future work could examine how end users experience dashboards and LLMs in practice, and how their expectations align with those of the experts we interviewed.
Revisiting participants over time would also allow us to examine which expectations translate into practice and which change as people gain experience with the tools.

\paragraph{Temporal and methodological limitations.}

The interviews combined descriptions of current practice with expectations about the future. They were conducted between November 2025 and August 2026, so participants' experiences reflect the tools and organizational practices they encountered during that period. Participants interviewed at different times may also have encountered different capabilities. We did not observe their work over time or independently evaluate the systems they described. The study therefore identifies how experts anticipate dashboard practices changing, but cannot establish whether these changes will occur or what their effects will be. Participants' descriptions of productivity and errors also reflect their experiences rather than independent measurements.
\section{Conclusion}
\label{sec:conclusion}

The promise of generating dashboards through AI raises questions about the future of the dashboard and the people working with it. Through interviews with 16 experts, we examined these questions in relation to current dashboard practices and participants' expectations. Almost all participants expected dashboards to persist, with recurring questions, monitoring, and established reporting practices giving people reasons to return to them. They also described possibilities for adaptive views and different interaction modalities. Language, graphical controls, and gestures could support different tasks, while interaction itself remained part of how users explore the data and develop their understanding.

Participants expected changes in the work behind these dashboards as well. Generating code and visualizations could leave authors with more work in specifying requirements, selecting suitable outputs, and checking their correctness. Understanding the audience, applying design knowledge, and communicating with stakeholders remained important in their expectations. Greater independence for end users also raised questions about whether they would have the knowledge to assess the dashboards they created. These concerns connect the future of dashboard interfaces with the responsibilities of those who create, use, and maintain them.

We developed seven opportunities for research and practice around validation, education, maintenance, organizational guidance, adaptation, novel visualizations, and accountability for AI-generated content. As dashboard creation becomes more accessible, research and industry will need to consider how people recognize mistakes, preserve shared understanding, sustain dashboards over time, and remain answerable for what those dashboards show. The value of future dashboards will depend on how well these responsibilities are supported alongside the ability to generate the dashboard itself.

%% ---------------------------------------------------------------------
%% Acknowledgments
%% ---------------------------------------------------------------------
\section*{Acknowledgments}
    This work was partially supported by Villum Investigator grant VL-54492 by Villum Fonden.
    Any opinions, findings, and conclusions expressed in this material are those of the authors and do not necessarily reflect the views of the funding agency.

\section*{Use of Generative AI}

The authors used generative AI tools: Gemini 3.8 Image Flash for generating the teaser figure. We also used Claude by Anthropic for generating the tables and support parts of the analysis. All outputs were reviewed and verified by the authors.

%% ---------------------------------------------------------------------
%% References
%% ---------------------------------------------------------------------

\bibliographystyle{ACM-Reference-Format}
\bibliography{main}
\end{document}